# Digital Footprints of Streaming Devices

Sundar Krishnan
Department of Computer Science
Angelo State University
skrishnan@angelo.edu

William Bradley Glisson
College Of Engineering and Science
Louisiana Tech University
glisson@latech.edu

## Abstract

*These days, there are many ways to watch streaming videos on television. When compared to a standalone smart television, streaming devices such as Roku and Amazon Fire Stick have a plethora of app selections. While these devices are platform agnostic and compatible with smartphones, they can still leave behind crumbs of sensitive data that can cause privacy, security, and forensic issues. In this paper, the authors conduct an experiment with streaming devices to ascertain digital footprints from network traffic and mobile forensics that they leave behind.*

## 1. Introduction

Streaming is a technology capable of transmitting continuous compressed data over the Internet in real-time instead of being saved to a storage drive before transmission [1]. Streaming data is also known as media that can be video, audio, closed captioning, ticker tape, or real-time text. For an end-user, streaming data is like live television broadcasts, which is processed almost instantaneously by a streaming device without waiting to download the file. Over the years, advances in computer networking, cheaper Cloud storage, growth of social media, and smartphones combined with powerful home computers, smart TVs, smartphone usage growth, and modern operating systems have made streaming of media faster, practical, and affordable for ordinary consumers.

In broadcasting jargon, live streaming is also known as over-the-top content (OTT). OTT is the audio, video, and other media content streamed over the Internet without control, management, or distribution of content by a Multiple System Operator (MSO). OTT video viewing is a commonplace media activity in the US, growing not only in reach but also in time spent, according to a ComScore report [2]. According to this report [2], about 82.4 million of US homes watched OTT videos in 2021, showing an annual increase of 22% from the previous year. While several forces such as smart TVs and 4G networks have driven the growth of OTT, a significant contribution has been the rapid growth of streaming devices, also known as “living room devices” or “OTT devices”. Households in the US watching OTT video content via streaming devices represents the largest group, with an average of 100 hours per month [2]. There is also a strong growing appeal for OTT in households with children vulnerable to security and privacy [2].

As of the third quarter of 2022, Roku and Amazon were the biggest market players [3] [4], continuing the trend set since 2018 [4], with each touting affordable devices that are packed full of desirable features and services. These devices are increasingly intelligent, flexible, and scalable to consumers’ needs, with the ability to integrate tightly with other products and online services. Smartphone Apps from television channels also interface with these streaming devices providing services. Often fast and reliable broadband Internet is a prerequisite for these devices for smooth streaming. These devices suggest program options to the users by relying on viewed/usage data, user preferences, and viewing/listening habits, allowing for feedback, and ratings.

Streaming media players can be considered an Internet-of-Things (IoT) appliance [5]. Every major smart TV streaming service logs the user’s watching habits. According to Amazon's privacy policy [6], the Firestick device collects data regarding user’s usage of the device and its capabilities, such as the navigation of the home screen and selection of device settings (such as device language, display size, Wi-Fi and Bluetooth choices). Roku’s privacy statement [7] states that they gather a wide range of information about the user, including the user’s name, email address, postal address, phone number, birth date, demographic data, location, device and usage information, and a ton of information about the content the user watches. According to Roku, they may also gather audio data when you utilize voice-activated capabilities, as well as data on your music, video, and photo files if you use one of their Roku Media Players to watch or listen to such content. IoT devices often use the same Wi-Fi network and generally trust other nodes on the network. However, this assumption can lead to accidental exposures as communication channels shared by other devices in the network can also be maliciously accessed by remote websites with just a small amount of manipulation. While streaming devices mostly encrypt their traffic, the network traffic behind a home router can be susceptible to malicious exploits such as sniffing, replay attacks, DNS rebinding attacks, session hijacking, digital identity leaks, and data leaks. In this paper, the authors assess network protocols used by Amazon and Roku streaming devices, the packet flow between their clients, their authentication management, their cipher suites used, their idle state connections, and examine any residual data they leave behind from a security and privacy perspective. The authors also examine the Smartphone devices used in tandem with these streaming devices to acquire residual and cached data that can infringe on the security and privacy of the users. A forensic footprint comparison between the two streaming devices is finally undertaken.

## 2. Background and Related Work

Existing literature has been reviewed in five sections; 1. IoT background to understand the prevalence of IoT devices in our surroundings, 2. Vulnerabilities in IoT Devices, 3. Privacy concerns in IoT Devices, 4. IoT Device Forensics, and 5. Security of IoT Devices.

### 2.1. Internet of Things (IoT) Background

According to Cisco, anywhere between 25 billion to 125 billion connected devices are expected to function worldwide by 2030 [8]. IoT refers to the billions of low-cost smart devices that connect to the Internet. Internet-of-things has evolved due to comingling of multiple technologies like real-time analysis and machine learning with traditional embedded systems. Per Hou et al. [9], IoT devices are constantly connected to the Internet and usually have default security configurations. Corser et al. [10] recommend device manufacturers limit the network traffic volume that IoT devices can generate to levels reasonably needed to perform their functions. Sivanathan et al. [11] attempt to classify IoT devices based on traffic characteristics as the traditional method of classification based on MAC address and DHCP negotiation is not always accurate with IoT devices. Unlike many other IoT devices, the network traffic pattern of streaming devices is meant to be continuous, and there is almost no sleep time between work cycles.

### 2.2. Vulnerabilities in IoT Devices

The National Institute of Standards and Technology (NIST) [12] defines a security vulnerability as “A weakness in the computational logic (e.g., code) found in software and hardware components that, when exploited, results in a negative impact to confidentiality, integrity, or availability”. Articles detailing vulnerabilities of Internet-facing Roku devices, and successful exploits, are often discussed on Shodan [13]. With Nessus as their vulnerability scanner, Stout et al. [14] use Shodan to identify vulnerabilities against IoT devices on the Internet. In their study, they conclude that a significant number of IoT devices are vulnerable to exploits. Angrishi [15] outlines the anatomy of IoT botnets, their operation, and their basic mode of attacks exploiting vulnerabilities. Per Angrishi [15] most of the IoT malware are Linux based, residing in RAM with limited or no side effects to the performance of the IoT device.

### 2.3. Privacy in IoT Devices

Security and privacy issues have been widely documented with IoT devices [16]. Nikas et al. [17] highlight the security and privacy of streaming services using Kodi as their testbed. In their study, the use of an unencrypted SQLLite database by Kodi can easily outline the various streaming services used, the last time accessed and how many times the service was used. In addition to various remediations, the authors suggest that the current model of remote services switch into using encrypted network traffic instead of HTTP. However, in an analysis of four commercially-available smart home IoT devices, Apthorpe et al. [18] conclude that despite traffic encryption, network traffic rates of all devices revealed user activities. Angrishi [15] concludes that the result of IoT evolution would bring about privacy-violating interactions between IoT devices, inventory attacks, information linkage and lifecycle transitions.

### 2.4. IoT Device Forensics

IoT Forensics pertains to the forensic acquisition of digital data from IoT devices. It is a branch of digital forensics covering the basic tenants of tamper-proof acquisition, preservation, and documentation. MacDermott et al. [19] discuss the need for a framework and process for performing IoT-based forensics. The authors also acknowledge that the main IoT forensics challenge is data acquisition by knowing exactly where the data is and acquiring it. They also highlight the gap in the field of Digital forensics, focusing on IoT devices. Nieto et al. [20] use the digital witness solution in their custom methodology to collect digital evidence balancing IoT forensics and privacy. Conti et al. [21] discuss forensic challenges in IoT environments, like attack attribution and evidence identification, collection, correlation, and preservation. Dorai et al. [22] examine Nest artifacts produced by an iPhone and develop a custom forensic tool to process Nest working data into a readable report for forensic analysis. Kebande et al. [23] compare existing forensic framework models and propose a digital forensic investigative framework that complies with ISO/IEC 27043:2015 international standards, and functions with an acceptable degree of certainty according to ISO/IEC standards.

### 2.4. Security of IoT Devices

As IoT devices grow in volume and acceptance, the security of services on their network and items they connect to the Internet are the focus of IoT security. Since IoT devices are mainly built towards functionality, embedded systems within these devices lack security features during their design phase. Using an IoT testbed for vulnerability scans, protocol evaluations, authentication, etc. Tekeoglu et al. [24] conclude that many IoT devices possess serious vulnerabilities, and the investigation of security and privacy on each device is complex and involves manual effort. Copos et al. [25] study two of the most popular smart home devices and conclude with reasonable accuracy that an attacker can detect if the home is occupied based on network traffic. Mollers et al. [26] analyze of HomeMatic automation devices and predict user behavior from wireless home automation communications. Stout et al. [14] survey various IoT domains for the shortcomings and challenges in securing IoT devices and their interactions with cloud and enterprise applications. Conti et al. [21] also highlight the security challenges of IoT environments such as authentication, privacy, secure architecture, authorization, and access control. In a staged experiment, Tekeoglu et al. [27] setup a Chromecast streaming device and studied the security and forensic aspects in its functioning. They conclude that data transmission patterns leak personal information outside of the home network raising privacy concerns.

As highlighted, security and forensic challenges abound with IoT devices. A literature gap exists in studying other popular streaming devices. In this paper, the authors focus on two popular streaming devices (Roku and Amazon Fire Stick) to understand their security readiness and the forensic footprint that they leave behind.

## 3. Test Environment and Design

The design focus of the test environment was to mimic a home network environment, commonly used functionality, and limit traffic noise from other devices on the home network. A product survey helped decide the combination of the devices and software for this design. The list of software tools used in this environment is listed in Table- 1, and Table-2 lists the hardware used.

| Item | Description |
|---|---|
| Smartphone Forensics | Paraben E3 DS [28] |
| Packet sniffer | Wireshark [29] |
| Packet analyzer | Network Miner [30], and Solarwinds Response Time Viewer [31] |
| Vulnerability scanner | Nessus Community Edition [32] and NMap [33] |

**Table 1 - Software Components of the Test Environment**

The software applications used were chosen due to their low-cost availability and functionality. All software excepting Paraben were free to download while few required product registrations by their vendors.

| Item | Description |
|---|---|
| Streaming Media Devices | Amazon Fire Stick with Alexa Remote (Quantity 1), Roku Express (Quantity 1) |
| Television (LCD Monitor) | Spectre 41 Inch flat Screen, TV with HDMI and USB input, ports (Quantity 1) |
| Smartphone | Alcatel SIM-enabled Android, Tablet, Model 9024W, Android OS version 7.1.1, 2GB RAM, Security patch, Nov-6 2017 (Quantity 1) |
| Network Switch | D-Link DGS 1100-10MP, smart network switch, (Quantity 1) |
| Wireless Access Point | Tenda Wireless router, Model W268R 150 Mbps, (Quantity 1) |
| Monitoring PC | Dell Optiplex 7010, Windows-10 OS, 16 GB RAM and 125 GB SSD, (Quantity 1) |

**Table 2 - Hardware Components of the Test Environment**

The Smartphone device used in this experiment was an unencrypted Android tablet with similar device functionality to that of a commonly used Android Smartphone. The network switch was configured for port mirroring and DHCP was turned off. A commonly found LCD television served the purpose of viewing streaming videos from these devices. The wireless access point was created by editing the Tenda router settings and enabling the access point option while disabling DHCP. The password for the router was set to a strong password as per NIST guidelines [34]. The streaming devices (Amazon Fire Stick and Roku Express) were connected to the HDMI port of the Television while drawing their power from the Television USB port.

To achieve port-mirroring on the network switch, its port # 1 was connected to the access point using a Cat5e Ethernet cable. The switch port #2 was connected to the motioning PC using a Cat5e Ethernet patch cable. Using the switch management web interface, switch port #1 traffic was mirrored to port #2. Thus, all traffic from switch port#1 will also flow into switch port#2. Switch port #3 was connected to an external gateway router connecting further to the Internet.

## 4. Methodology

In this section, the authors describe the laboratory environment used to examine the streaming devices. The goal of this experiment was to put the devices to use and then determine if it was possible to retrieve any residual data of interest. The network topology diagram is shown in Figure- 1. To verify the research questions of this study, the following hypotheses were formulated.

**H1**: Forensic data of interest from an IoT device can be recovered by sniffing its network traffic.

**H2**: Residual data recovered from the Smartphone used to remote control the IoT device can lead to determining IoT device/user credentials, network connection logs, and user data.

The experiment was divided into two stages: one for the Amazon Fire Stick and the other for Roku Express. During each stage, the following methodology was followed in sequence.

### 4.1 Methodology for Amazon Fire Stick and Roku Express

The methodology for the experiment consisted of the below seven steps repeated for each streaming device. Care was taken to document findings in a concise manner.

**Step 1** - Evaluate the streaming device
Examine the new out-of-the-box device without root and refer to the product manual for proper usage.

**Step 2** - Environment preparation
A Tenda Wireless router (W268R 150 Mbps) was used as a wireless access point and was connected to port#1 of the network switch. The Streaming device and Android device (tablet) were connected to this access point over Wi-Fi. The appropriate App for the streaming device was downloaded from the Android Playstore onto the Android device. Wireshark, Network Miner, Paraben, and Solarwinds ”Response Time Viewer for Wireshark” software was downloaded from the Internet and installed on the monitor PC. Completed streaming device registration with vendor website as part of the initial setup process. Angry IP Scanner [35] was used to scan and identify all IP addresses on the network topology.

**Step 3** - Activate and register streaming devices
Start a new Wireshark capture on the monitor PC to monitor mirrored port traffic on switch port#1. Connect the Amazon Fire TV Stick to the Television and power it on. Follow the instructions on the TV screen to register the device and setup Wi-Fi connectivity to the Tenda Access Point (connected to the layer-3 switch and having its port traffic mirrored). Stop Wireshark capture and save capture traffic data as “streaming_device_name_registration.pcap” file for later analysis. Keep the Amazon Fire Stick and television powered on. Gather the IP address of the streaming device.

**Step 4** - Stream video on the device
Start a new Wireshark capture on the monitor PC to monitor the mirrored port traffic on switch port#1. Keep the device paired with the Tenda Wi-Fi access point that’s connected to the layer-3 switch and have its port traffic continuously mirrored. Using the Amazon Fire TV Stick provided remote, select a video for streaming while the device is connected to the Television. After playing two videos each for 5 minutes, stop Wireshark and save network traffic captured as “streaming_device_name_usage.pcap” file for later analysis. Keep the streaming device and television powered on.

**Step 5** - Idle state of streaming media device

Start a new Wireshark capture on the monitor PC to monitor the mirrored port traffic on switch Port#1. Capture Amazon Fire Stick traffic when the device is idle for 5 minutes. The Television screen will switch to screensavers during this time. Stop Wireshark capture and save captured network data as “streaming_device_name_IdleState.pcap” file for later analysis. Leave the streaming device and television powered on.

**Step 6** - Device remote via Android App

Start a new Wireshark capture on the monitor PC to monitor the mirrored port traffic on switch Port#1. On the Android tablet, download and install remote device App from the Android Playstore. Open the App on the tablet and perform device control and video streaming for 5 minutes. Stop Wireshark capture and save captured data as “streaming_device_name_remote_usage.pcap” file for later analysis. Leave the streaming device and Television powered on.

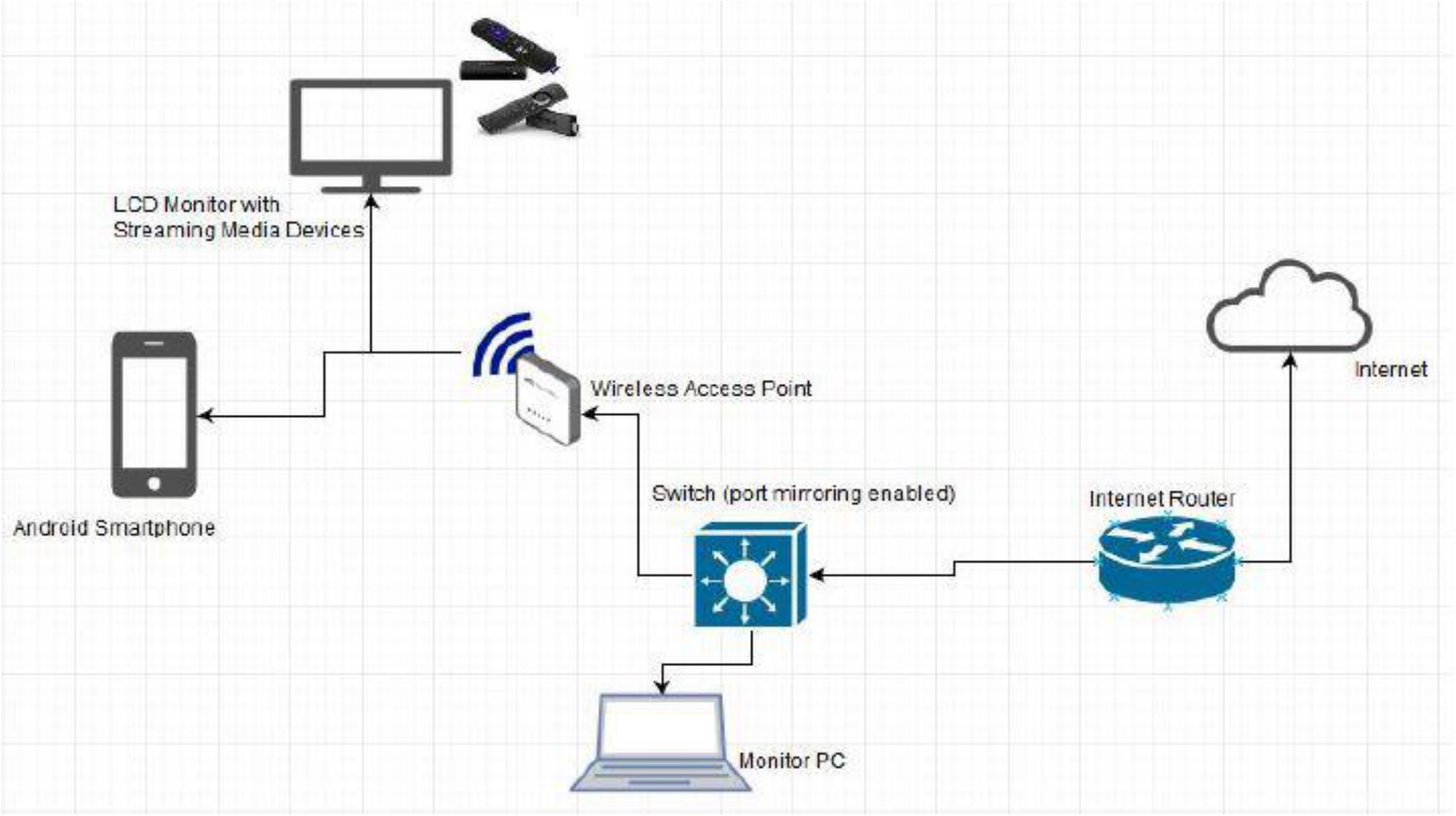

**Figure 1: Network Topology**

**Step 7** - Vulnerability assessment

Nmap- From the monitoring PC, start a Nmap scan against the streaming device for vulnerabilities. Save the scan results for later analysis. Use the below commands for the scans;

nmap -sS sU -T4 -A -v 192.168.x.x
nmap -p 1-65535 -T4 -A -v 192.168.x.x

Nessus - Start a new Nessus scan. Post scan completion, save scan results within Nessus for later analysis. Leave the steaming device and Television powered on.

**Step 8** - Power off the Device

Start a new Wireshark capture on the monitor PC to monitor the mirrored port traffic on switch Port#1. First, power off the streaming device and then power the television down. Stop Wireshark capture and save capture data as “streaming_device_name_shutdown.pcap” file for later analysis.

**Step 9** - Forensics

Using Paraben, start a new forensic case and conduct Android device forensics after the acquisition of smartphone device without rooting. Save results for later analysis.

**Step 10** - Prepare for Analysis

Gather all the data in files collected for a systematic analysis of the traffic captured from Wireshark, vulnerabilities from Nessus and Smartphone acquisition.

## 5 Analysis and Results

The choice of the network access point throughput did impact the quality of the streaming service to a certain degree but with time, the streaming devices performed better due to buffering.

**Figure 2: Sankey Diagram of traffic flow and packets during Roku Express registration (L) and use(R)**

### 5.1 Network Traffic Analysis

Wireshark and Network Miner were primarily used to analyze the five .pcap files obtained when following the above methodology.

- streaming_device_name_registration.pcap
- streaming_device_name_usage.pcap
- streaming_device_name_IdleState.pcap
- streaming_device_name_remote_usage.pcap
- streaming_device_name_shutdown.pcap,

Each .pcap file was analyzed for traffic flow to domains.

**Amazon Fire Stick**- After filtering streaming_device_name_registration.pcap, pcap (FireStick_registration.pcap) Wireshark capture for Fire Stick (source) IP, for 11-minute traffic analysis, the Fire Stick device connected with over 110 unique IPs. It was noisy on the network as it had more data transmission and traffic than the Roku device. Most of the traffic was encrypted. However, compared to the Roku device, much more unencrypted image files were recovered from this traffic after reassembly, that seems to be streaming channel icons. The image size was small and served for surfed program thumbnails. The bulk of the traffic was directed to Cloudfront and Amazon AWS servers.

After filtering streaming_device_name_usage.pcap (Firestick_usage.pcap) Wireshark for Fire Stick (source) IP, for a 9-minute traffic analysis, the Fire Stick device connected with over 70 unique IPs. Most of the traffic was encrypted.

Hundreds of unencrypted image files of small sizes were recovered from this traffic after reassembly that served as thumbnails for streaming channels or surfed programs. This was much more than the Roku device when channel surfing or streaming. The bulk of the traffic was directed to Cloudfront and amazon AWS servers.

Multiple .mp4 files (broken in parts) could be reassembled from the traffic dump (over TCP 80). Streaming media with clear audio could be replayed from a few files successfully. The other .mp4 files were encoded to protect digital rights concerns.

Other .pcap files generally contained encrypted traffic. No user/network credentials were exposed in plain text. Multiple security certificates were also found in the traffic.

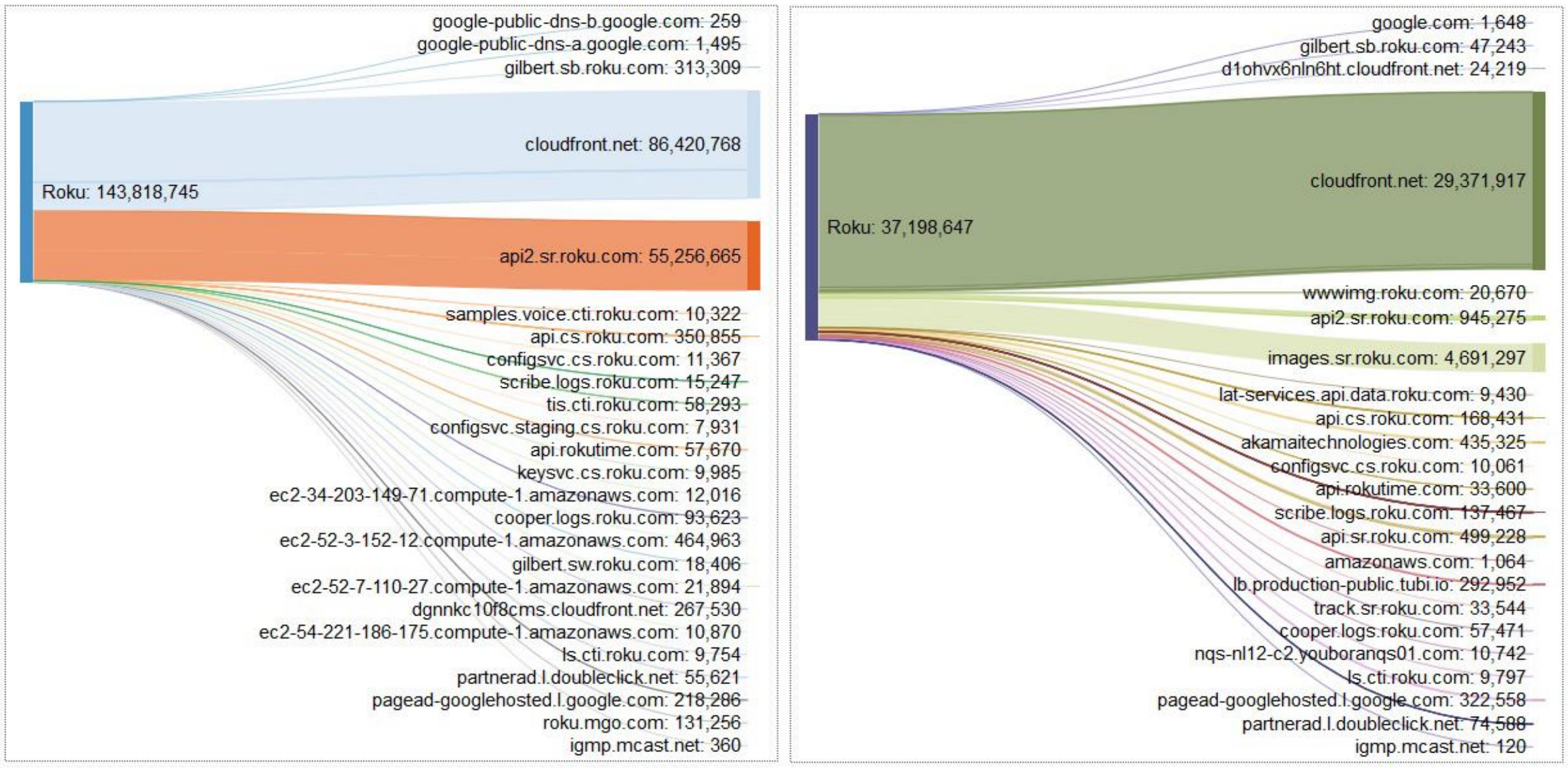

**Roku** – After filtering streaming_device_name_registration.pcap (Roku_registration.pcap), for Roku (source) IP, for 14-minute traffic analysis, the Roku device connected with over 70 unique IPs. Most of the traffic was encrypted. A few unencrypted image files were recovered from this traffic after reassembly, which seems to be that of various streaming channels and programs. The image size was small (Kb) and probably served for thumbnails. Traffic to cooper.logs.roku.com and digdugdata.roku.com was abundant and the domain name suggests some level of active logging. Traffic to configsvc.staging.cs.roku.com could be interpreted as device registration data. The bulk of the traffic was directed to Cloudfront and Amazon AWS servers.

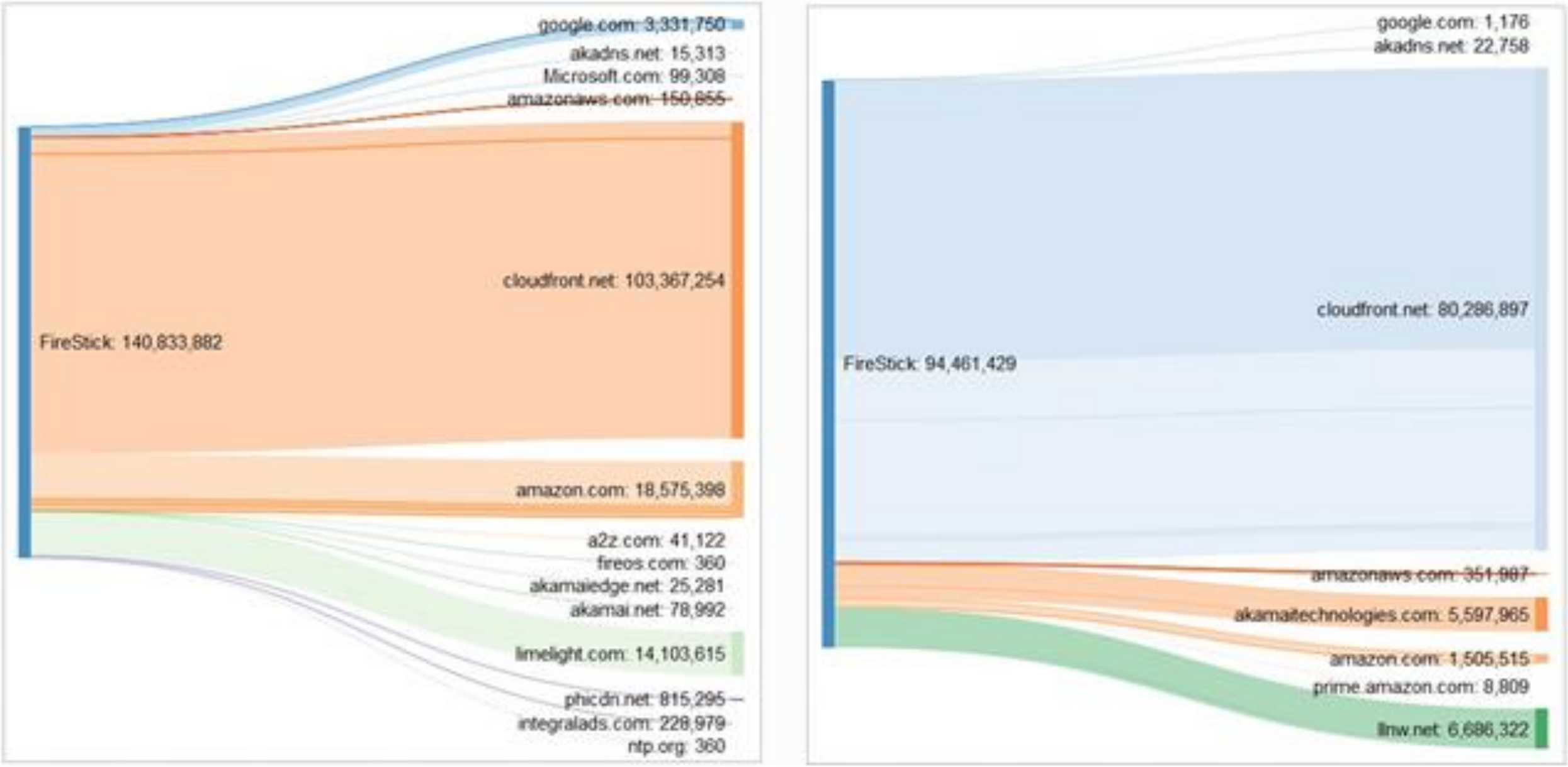

**Figure 3: Sankey diagram of traffic flow and packets during Fire Stick Registration (L) and use(R)**

After filtering streaming_device_name_usage.pcap (Roku_usage.pcap), for Roku (source) IP, a 3-minute traffic analysis was done, and the Roku device connected with over 75 unique IPs. Most of the traffic was encrypted. Few image files were recovered from this traffic after reassembly that did seem to be snapshots of the program being watched. The image size was small and probably served for thumbnails. Traffic to Scribe.logs.roku.com was found in abundance, and the domain name would suggest some sort of active logging. The bulk of the traffic was directed to Cloudfront servers. Cloud providers Cloudfront and Amazon AWS hosted most of the target endpoints for the device. Other .pcap files of Roku generally contained encrypted traffic and plenty of security certificates.

### 5.2 Vulnerability scans

Vulnerability scans for both the streaming devices were performed using Nmap and Nessus. Scan highlights are for Nmap in Table-3 and for Nessus in Table-4.

| Fire Stick | Roku |
|---|---|
| nmap -sS sU -T4 -A -v 192.168.x.x and<br>nmap -p 1-65535 -T4 -A -v 192.168.x.x | nmap -sS sU -T4 -A -v 192.168.x.x and<br>nmap -p 1-65535 -T4 -A -v 192.168.x.x |
| Identified OS as a variant of Linux 3.x | Identified OS as a variant of Linux 3.x |
| TCP open ports 8009, 40289, 55763, 60000 with UDP open ports 67, 137, 162, 1066, 1080, 1900, 5353, 8001, 20389, 22124, 31365, 35438, 49166, 49171 and 65024 | TCP open ports 8060, 53790 with UDP open ports 1900 |
| TCP Port 8009 allowed http service with Amazon Whiserplay DIAL REST service | |

**Table 3 – Nmap scan highlights**

| Fire Stick | Roku |
|---|---|
| A polite scan was configured without credentials | A polite scan was configured without credentials |
| Identified device OS as a Linux Kernel 2.6 | Identified device OS as a Linux Kernel 2.6 |
| Scan resulted in only one vulnerability that was listed informational with zero risk. | Multiple Ethernet Driver Frame Padding Information Disclosure (Etherleak)- CVSS Base Score: 3.3 was listed as a low risk vulnerability. 15 other vulnerabilities were listed as informational with zero risk.<br>TCP port 8060 was listed as running a web server and the endpoint was a UPnP client. http://192.168.x.x:8060/ yielded deviceType, serialNumber etc. as in Figure-5 |

**Table 4 – Nessus scan highlights**

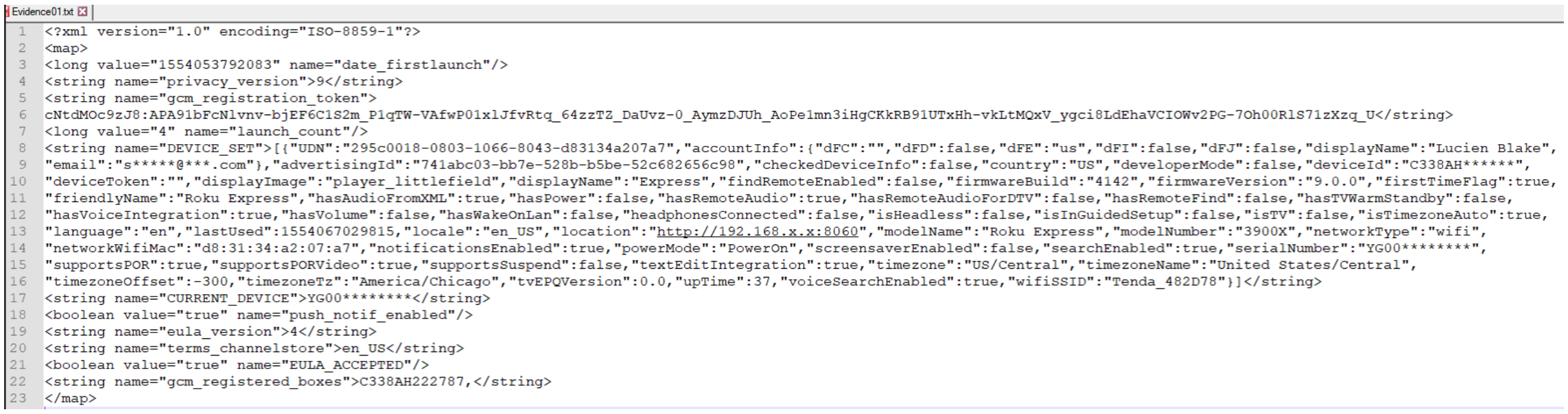

```
<?xml version="1.0" encoding="ISO-8859-1"?>
<map>
<long value="1554053792083" name="date_firstlaunch"/>
<string name="privacy_version">9</string>
<string name="gcm_registration_token">
cNtdMOc9zJ8:APA91bFcNlvnv-bjEF6C1S2m_P1qTW-VAfwP01xlJfvRtq_64zzTZ_DaUvz-0_AymzDJUh_AoPe1mn3iHgCKkRB91UTxHh-vkLtMQxV_ygci8LdEhaVCIOWv2PG-7Oh00RlS71zXzq_U</string>
<long value="4" name="launch_count"/>
<string name="DEVICE_SET">[{"UDN":"295c0018-0803-1066-8043-d83134a207a7","accountInfo":{"dFC":"","dFD":false,"dFE":"us","dFI":false,"dFJ":false,"displayName":"Lucien Blake",
"email":"s*****@***.com"},"advertisingId":"741abc03-bb7e-528b-b5be-52c682656c98","checkedDeviceInfo":false,"country":"US","developerMode":false,"deviceId":"C338AH******",
"deviceToken":"","displayImage":"player_littlefield","displayName":"Express","findRemoteEnabled":false,"firmwareBuild":"4142","firmwareVersion":"9.0.0","firstTimeFlag":true,
"friendlyName":"Roku Express","hasAudioFromXML":true,"hasPower":false,"hasRemoteAudio":true,"hasRemoteAudioForDTV":false,"hasRemoteFind":false,"hasTVWarmStandby":false,
"hasVoiceIntegration":true,"hasVolume":false,"hasWakeOnLan":false,"headphonesConnected":false,"isHeadless":false,"isInGuidedSetup":false,"isTV":false,"isTimezoneAuto":true,
"language":"en","lastUsed":1554067029815,"locale":"en_US","location":"http://192.168.x.x:8060","modelName":"Roku Express","modelNumber":"3900X","networkType":"wifi",
"networkWifiMac":"d8:31:34:a2:07:a7","notificationsEnabled":true,"powerMode":"PowerOn","screensaverEnabled":false,"searchEnabled":true,"serialNumber":"YG00********",
"supportsPOR":true,"supportsPORVideo":true,"supportsSuspend":false,"textEditIntegration":true,"timezone":"US/Central","timezoneName":"United States/Central",
"timezoneOffset":-300,"timezoneTz":"America/Chicago","tvEPQVersion":0.0,"upTime":37,"voiceSearchEnabled":true,"wifiSSID":"Tenda_482D78"}]</string>
<string name="CURRENT_DEVICE">YG00********</string>
<boolean value="true" name="push_notif_enabled"/>
<string name="eula_version">4</string>
<string name="terms_channelstore">en_US</string>
<boolean value="true" name="EULA_ACCEPTED"/>
<string name="gcm_registered_boxes">C338AH222787,</string>
</map>
```

**Figure 4: Android Forensics – contents of Roku com.roku.remote_preferences.xml**

```
Here is a summary of http://192.168.[redacted]:8060/ :

deviceType: urn:roku-com:device:player:1-0
friendlyName: Express
manufacturer: Roku
manufacturerURL: http://www.roku.com/
modelName: Roku Express
modelDescription: Roku Streaming Player Network Media
modelName: Roku Express
modelNumber: 3900X
modelURL: http://www.roku.com/
serialNumber: YG00H2[redacted]
ServiceID: urn:roku-com:serviceId:ecp1-0
        serviceType: urn:roku-com:service:ecp:1
        controlURL:
        eventSubURL:
        SCPDURL: ecp_SCPD.xml
ServiceID: urn:dial-multiscreen-org:serviceId:dial1-0
        serviceType: urn:dial-multiscreen-org:service:dial:1
        controlURL:
        eventSubURL:
        SCPDURL: dial SCPD.xml
less...
```

**Figure 5: Roku webpage displaying metadata**

From the above, it can be deduced that the mere running of a webserver by the Roku device and the availabity of the website over HTTP can yield metadata that could be hidden.

## 5.3 Android Forensics

Android forensics was primarily conducted using Paraben. Since the android device (tablet) was not rooted, only partial data extraction was possible. While not a whole lot of data was recovered from Fire Stick App, data acquired from the Roku App showed network and user metadata. Roku had a webpage that displayed detailed user registration, IP address, and wireless connection details. Details of the data acquired as in Table -5 and Figure-4

| **Fire Stick** | **Roku** |
|---|---|
| PERMISSIONS FOR Fire TV<br><br>**Highly Suspect:<br>android.permission.AUTHENTICATE_ACCOUNTS<br>android.permission.CHANGE_NETWORK_STATE<br>android.permission.CHANGE_WIFI_MULTICAST_STATE<br>android.permission.MANAGE_ACCOUNTS<br><br>*Suspect:<br>android.permission.ACCESS_FINE_LOCATION<br>android.permission.RECORD_AUDIO<br>android.permission.USE_CREDENTIALS<br>android.permission.WRITE_EXTERNAL_STORAGE<br>android.permission.ACCESS_COARSE_LOCATION<br>android.permission.CHANGE_WIFI_STATE<br>android.permission.GET_ACCOUNTS<br>android.permission.ACCESS_NETWORK_STATE<br>android.permission.ACCESS_WIFI_STATE<br><br>Low Suspect:<br>android.permission.INTERNET<br>android.permission.READ_EXTERNAL_STORAGE<br>android.permission.READ_PHONE_STATE<br>android.permission.VIBRATE<br>android.permission.WAKE_LOCK | com.roku.remote_preferences.xml – detailed the registered user, country, email addresswi-fi network details, IP address etc. as in Figure-4. |

**Table 5 – Android forensics highlights**

## 6. Conclusion

As users increasingly embrace electronic devices for convenience or entertainment, these devices leave behind footprints of forensic interest, leading to security and privacy concerns. When compared to Firestick on privacy, Roku is like the prying, gossipy neighbor as it keeps tabs on pretty much everything that the user does and later distributes this information to a huge number of marketers, channel providers, business partners, and others. While privacy can be controled on the device setting to certain degree, the security vulnerabilities of these devices lurk in the background oblivious to the user. In this experiment, the authors perform forensics of Amazon Fire Stick and Roku Express and scan them for vulnerabilities. The authors find that during their normal usage, forensic data could be acquired from their network traffic and the Android smartphone device used to control them. The recovery of thumbnails of programs surfed from both the devices, network and user metadata on Roku App, the Roku webpage displaying device metadata, and the successful network traffic reconstruction of .mp4 files from Fire Stick proves the two initial hypotheses by the authors. This experiment shows that with all the encryption safeguards used in today's age, plaintext data on devices can still be found as digital breadcrumbs, especially when using digital devices connected to the Internet.